# WHY IS SO HARD TO GROW A PERFECT CRYSTAL? – 2D AND 3D MONTE CARLO SIMULATIONS

**Bogdan Ranguelov, Alexander Karamanov**

Institute of Physical Chemistry, Bulgarian Academy of Sciences, Acad. G. Bonchev Str., bl. 11, Sofia – 1113, E-mail: rangelov@ipc.bas.bg

**ABSTRACT.** We focus our attention on Monte Carlo simulations of crystallization, which is one of the most important processes occurring in nature and technology of materials. Special attention is paid to the crystallization shrinkage and its consequences onto the growth of the new crystalline phase. We show that crystallization shrinkage stops after reaching the percolation threshold and that this is not the final stage of the crystallization process. Based on experimental evidences on sintering of diopside-albite systems and computer simulations we argue that crystallization continues inside the grains where the remaining liquid melt is a subject of accumulation of stress and because the stress relaxation by change in the volume is forbidden due to the already formed rigid core, the so called crystallization induced porosity appears. We show that even a relatively simple short range potential as the Lennard-Jones one, reveals such a complicated phenomenon as the crystallization induced porosity, which has a great importance onto the mechanical properties of the newly crystallized material.

## INTRODUCTION AND OVERVIEW

Crystallization is a process of formation of a new phase and it is of great importance not only in nature, but also in the technology of materials. In nature, one simple and typical example for crystallization is the formation of snowflakes and formation of rocks. Sugar and chocolate are also result of crystallization processes. In recycling industry, crystallization (and re-crystallization) is a concomitant process during sintering, which is a key method for obtaining new materials. In principal, the formation of a new phase by crystallization is always accompanied by volume shrinkage of the initial phase, herein and after considered to be a liquid monoatomic melt. The degree of this volume shrinkage mainly depends on the ratio of the densities of the initial and the newly formed phase. It also depends on the kinetics of formation of the new phase and the position of formation of over critical size nuclei onto which the growth process takes place. This growth process consumes the parent phase in a concurrent manner – bigger is the newly formed phase, bigger is the consumption from the parent phase. Thus, the new phase forms individual crystallites that are more or less miss oriented one to another and as a result they might form a rigid skeleton (of the new phase) which hinders the further volume shrinkage of the initial system. This stage of the system evolution is called percolation threshold. Depending on the nature of the system, the percolation threshold appears at different degrees of transformation towards the new phase. For "pure" systems like metals this threshold is observed after 60 – 70 % of a new phase formation, while for diopside-albite systems under sintering conditions the shrinkage becomes negligible after comparatively small amount – 10 – 20 % of crystal formation (Karamanov A., Pelino M., 2006). As a consequence after this moment the system is not able to decrease its volume and considerable amount of parent phase is locked into the rigid skeleton of the new phase. Since the system trend is to lower its energy, the crystallization process continues inside the formed rigid core regions. Thus the remaining liquid melt becomes a subject of stress accumulation, which leads to a void formation into the parent phase, namely a crystallization induced porosity (CIP). Strikingly, CIP improves the mechanical properties of the sintered materials (Karamanov A., Pelino M., 2008) and rises question about the role of such "defects", which are classically considered to decrease materials functionality. This phenomenon is the basic motivation to perform Monte Carlo simulations (MCS) on crystallization of liquid melts, comparing two cases – free standing melt and a melt confined into an already formed crystalline core. The main questions that we try to answer by means of MCS are: where starts the nucleation of the new phase – bulk or surface crystallization is the fastest process. What is the influence of the already formed rigid core on the crystallization of the residual parent phase, and where appears the CIP. In order to answer these questions we start our simulations on a free standing liquid melt, where we vary the size of the system and initial packing fraction, then we continue with a melt confined into a rigid core and finally we compare these two cases in 2D and 3D.

## MODEL AND ANALYSIS TECHNIQUES

We perform MCS on a simplest system of a monatomic liquid melt (one component system) using classical pair wise Lennard-Jones (LJ) potential in 2D and 3D for two different cases – free liquid melt and a liquid melt confined in a rigid core of an already formed crystalline phase. The interaction potential between two particles ($i$ and $j$) is defined as

$$V(r_{ij}) = 4\varepsilon\left[\left(\frac{\sigma}{r_{ij}}\right)^{12} - \left(\frac{\sigma}{r_{ij}}\right)^{6}\right] \qquad (1)$$

where $r_{ij}$ is the distance between the two particles, $\varepsilon$ is the depth of the potential well (interaction strength), and

$\sigma$ is the minimal possible distance at which the two particles repel each other like hard spheres. This is a pair wise potential with one global minimum and it inevitably leads to a formation of a hexagonal closed packed (*hcp*) crystalline structure. We use the NVT Metropolis sampling technique (Landau D., Binder K., 2009) – constant number of particles, constant volume and constant temperature and our simulations are performed according to the phase diagram of the LJ model (Tang, J. et al., 2002), with reduced temperature $T^* = k_B T / \varepsilon$, where $k_B$ is the Boltzman constant. The crystallization process is monitored at every Monte Carlo step by means of several quantities – the total energy of the system, the number of crystalline particles, the pair distribution function (PDF) and the histogram of particles displacements vs. their initial positions. The first minimum of PDF at each Monte Carlo step is used to determine the radius of the so called first coordination shell, which is used to determine the coordination number of each particle (degree of crystallinity). Here we use the strictest condition and in 2D case we label a particle as a crystal one only if it has exactly six neighbors into the first coordination shell. In the case of 3D we use the condition that a particle is labeled as a crystalline one if it has 10 or more neighbors into the first coordination shell. In all figures this crystal particles are colored in red, while the remaining amorphous or liquid particles are colored in cyan.

## SIMULATION RESULTS AND DISSCUSION

Our simulations are performed on an NVT ensemble using standard Metropolis criterion algorithm for accepting or rejecting the trial moves of the particles (Liu J., 2008) with a probability given by

$$p = \min\left[1, \exp\left(\frac{-\Delta E}{k_B T}\right)\right] \quad (2)$$

In all simulations we use a cut-off distance value of $6\sigma$ and a standard potential depth of $\varepsilon = 1$.

### Case1: Free standing melt

We start our simulations with a free standing melt of a relatively small system consisting of 400 particles with an initial packing fraction (PF) of value 0.72 at a reduced temperature 0.12. The result is presented on Figure 1, which consists of two columns: the left column represents the crystallization process of the melt (crystalline particles are labeled as red), the right column represent the same process, but of different point of view, namely the displacement of the individual particles, which are labeled as follows: if a particle displacement is lower than $2\sigma$, then this particle is colored in cyan, if the displacement is greater than $2\sigma$, then the corresponding particle is colored in red. Thus, we "monitor" simultaneously both the crystallization of the melt turning into a rigid new phase increasing the coordination number of each particle and the individual displacement of the particles, which is a result of the sucking action due to the already formed crystallites of the new phase. The two vertical black lines are guide to the eye towards the volume shrinkage of the whole system. Time scale runs from top to bottom, final images are obtained after ~ $10^6$ Monte Carlo steps.

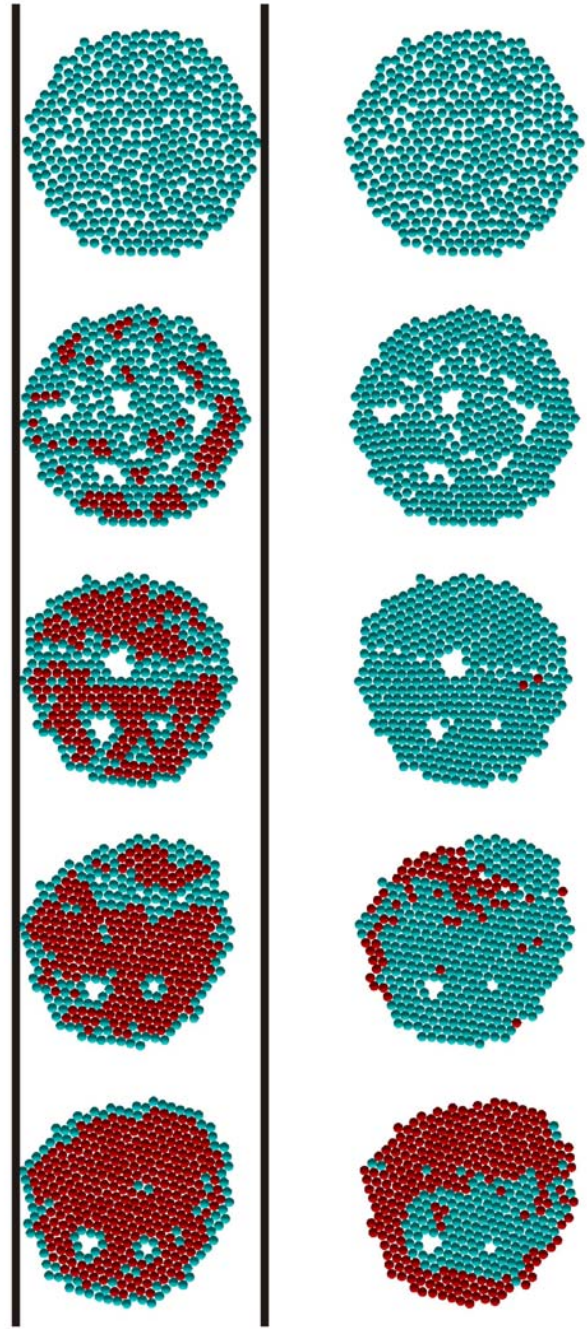

**Figure 1. 2D crystallization of 400 particles (free melt), PF = 0.72, T\* = 0.12; Left – coloring according to the coordination number of each particle; Right – coloring according to each particle displacement against its initial position.**

Figure 1 represents the initial milestone in our work – the system crystallizes by forming several overcritical nuclei, which consume the initial liquid melt, transforming it in a new crystalline denser phase. The initial system decomposes into simultaneously growing number of crystallites, which are generally speaking miss-oriented one by another, causing significant amount of defects into the system. Since the considered system is not too big and its borders are free, we observe clear shrinkage leading to a minimizing in the total energy and gradually healing of the defects, which are pushed toward the system boundaries. Thus, the final stage of the crystallization process is an almost perfect newly formed crystal phase with a minimal amount of so called point defects. These point defects are inevitable consequence of the crystallization process and generally are not a subject to be overcome – the system that crystallizes must pay a very high energetic cost in order to get rid of them. The right column of Figure 1 represents clear evidence that the shrinkage of the system proceeds by a movement of the outer particles toward the center, thus favoring the surface crystallization. Crystallization process of the whole system is very well observed by means of PDF function, shown on Figure 2.

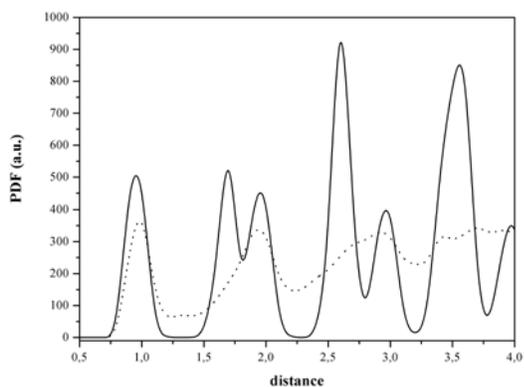

**Figure 2. Pair Distribution Function (PDF) of initial configuration (dotted line) and PDF after $10^5$ Monte Carlo simulation steps (solid line), simulation parameters are same as in Figure 1.**

Figure 2 represents a typical PDF for an amorphous liquid melt state in the beginning of our simulation (dotted line) with one defined peak at distance 1.0, and weak peak at distance 2.0. This presents the close range order in the liquid melt state. As the system tends to crystallize, long range order begins to appear (solid line), which is seen by the new formed peaks after $10^5$ Monte Carlo simulation steps.
Now we will double the size of the simulation system, preserving other parameters of the simulation the same. Initial and final results are shown on Figure 3.

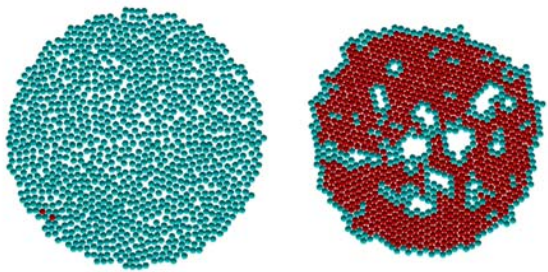

**Figure 3. Initial and final configuration of a larger system of more than 1000 particles. Colouring according to the coordination number.**

Figure 3 represents our results concerning a larger simulation system – it is clearly seen that the final state (after $10^6$ Monte Carlo steps) accumulates more defects and these defects are not individual point defects like in the smaller system, but voids in the final crystalline structure. The reason for this voids is that the larger system "produces" much more initial defects which during the crystallization process can coalescence in larger voids, and if they are far from the system boundary, their elimination is either impossible (thermodynamically unfavourable) or is very much time consuming. Elimination of an already formed defects is not easy, especially if the defects are in the form of point (singular) defects, or bigger voids. Relatively easy is the healing of grain boundaries between individual crystallites – see the movement of the slippery grain boundary on left column of Figure 1. On Figure 4 we show another characteristic feature of crystallization of a larger system. We monitor the volume shrinkage of the system (mean outer radius) and the degree of crystallinity vs. time.

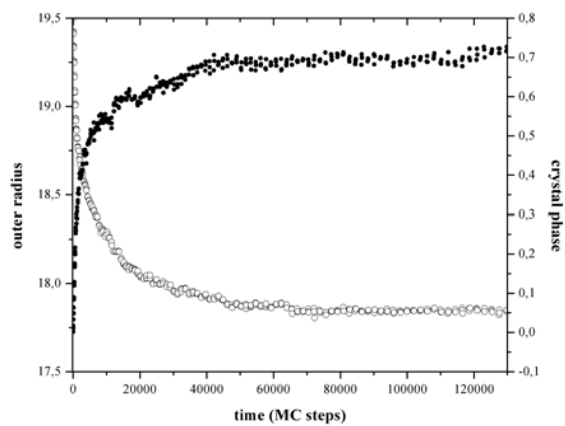

**Figure 4. Time dependence of the mean outer radius of the system (open circles) and the degree of crystallinity (solid circles).**

Figure 4 shows that the system shrinkage almost stops after developing a certain degree of a crystalline phase (percolation threshold around 70% of crystal phase), but not all of the initial system is fully crystallized – slow upper trend in crystallization is observed – solid circles in Figure 4. This means that after reaching the percolation threshold some amount of liquid melt remains inside the rigid "core" formed from the growing crystallites of the new phase.
Since the real crystallization process takes place in 3D, we continue our simulations with a free standing 3D melt. The result is shown on Figure 5, where we plot the half of the system - only particles that have positive value of z-coordinates are shown (left hand side coordination system is used). This gives us an opportunity to "look inside" the crystallizing system. The system consists of 3000 particles, initial PF is 0.5 and reduced temperature is 0.10. Colouring is according to the coordination number of each particle – crystalline particles are red. On Figure 5 we show the early stages of crystallization, the final image is obtained after only 1000 Monte Carlo steps. It is clearly seen, that the system starts to crystallize by forming overcritical nuclei predominantly into the surface layer of the system. This result is to be expected, since the surface particles have neighbours only from the inner volume of the system, and thus the crystallization shrinkage leads to an initial densification into the surface layer. Of course, crystallization is developed simultaneously inside the volume, and overcritical nuclei are found around the centre of the system too. Here we have to stress, that we use a very simple classical LJ potential, which is a very short range potential with one strong minimum, and that this potential is applicable mainly for monoatomic systems of noble gases or pure metallic systems.

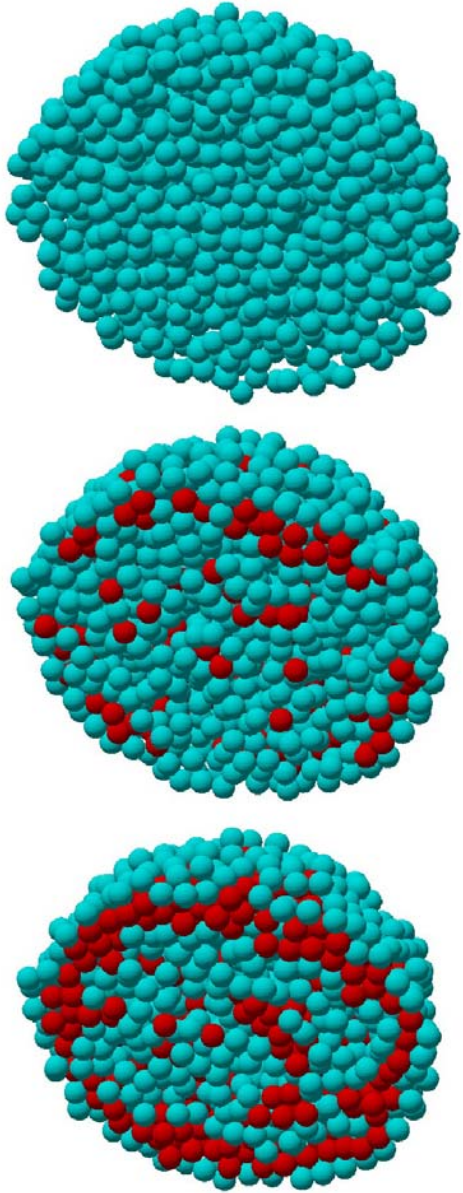

**Figure 5. 3D crystallization. PF = 0.50, T\* = 0.10. Coloring according to the coordination number of each particle. Only half of the system is shown (particles with positive value of z-coordinates).**

LJ potential is a basic potential used for simulations in condensed matter and we use it towards shedding light onto the basic physical phenomena of crystallization. More accurate, but complicated potentials that might be used are Morse potential, or Lennard-Jones-Gauss potential (Mizuguchi T., Odagaki T., 2009).

Since the overcritical nuclei are formed predominantly in the surface layer (at least in the initial stage of crystallization) they continue to grow and finally they form a rigid dense spherical core around the system. This is the first step of the whole crystallization process.

Careful examination of the rigid crystalline core (by means of "tomography like" slices) shows that it is formed all over the sub-surface layer of the system. Immediately after the formation of this core (reaching of a percolation threshold), the system stops its shrinkage, leaving a considerable amount of a liquid melt phase inside it. This amount of liquid phase tends to crystallize further, but since the system had already reached its percolation threshold and cannot further shrink its volume, we believe that a stress is accumulated into the liquid phase. This stress leads to a break into the liquid phase – this is the so called crystallization induced porosity, which forms closed pores inside the volume of the newly formed phase. This result is the second milestone in our work, and it leads to further simulations concerning a confined liquid melt in an already formed rigid crystalline core.

**Case2: Confined melt in a rigid crystalline core**

In contrary of the previous Case1, now we will put a liquid melt inside of an already formed crystalline core of a maximum possible density (*hcp* crystalline structure) – see Figure 6. After thermal equilibration of the liquid melt, the whole system is left to crystallize at a given reduced temperature, and the monitor scheme repeats that from Case1. Only the outermost two layers of the crystalline phase are fixed.

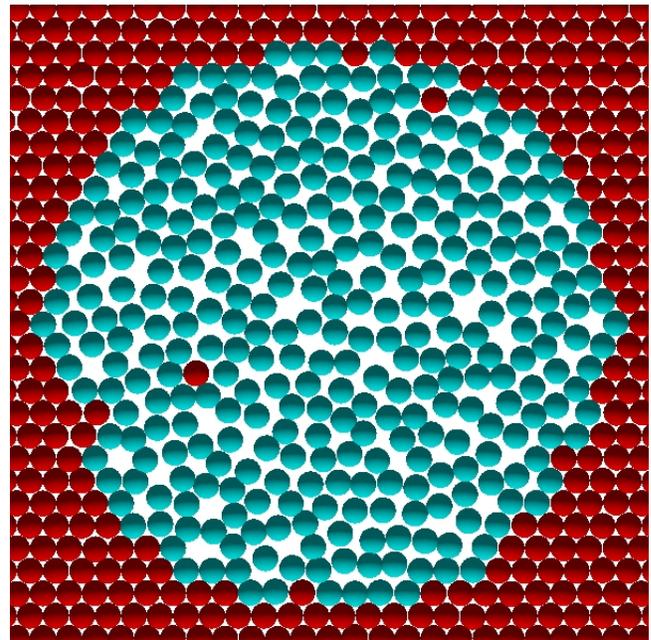

**Figure 6. 2D initial configuration – liquid melt inside an already formed crystalline core. Crystal (*hcp*) particles labelled as red.**

Result form our simulation is shown on Figure 7, where we show a similar presentation as on Figure 1, but for Case2. Again left column presents coloring according to the coordination number of each particle, while the right column presents the indication of individual particle movement. Last images (bottom) on Figure 7 are obtained after $10^5$ MC steps.

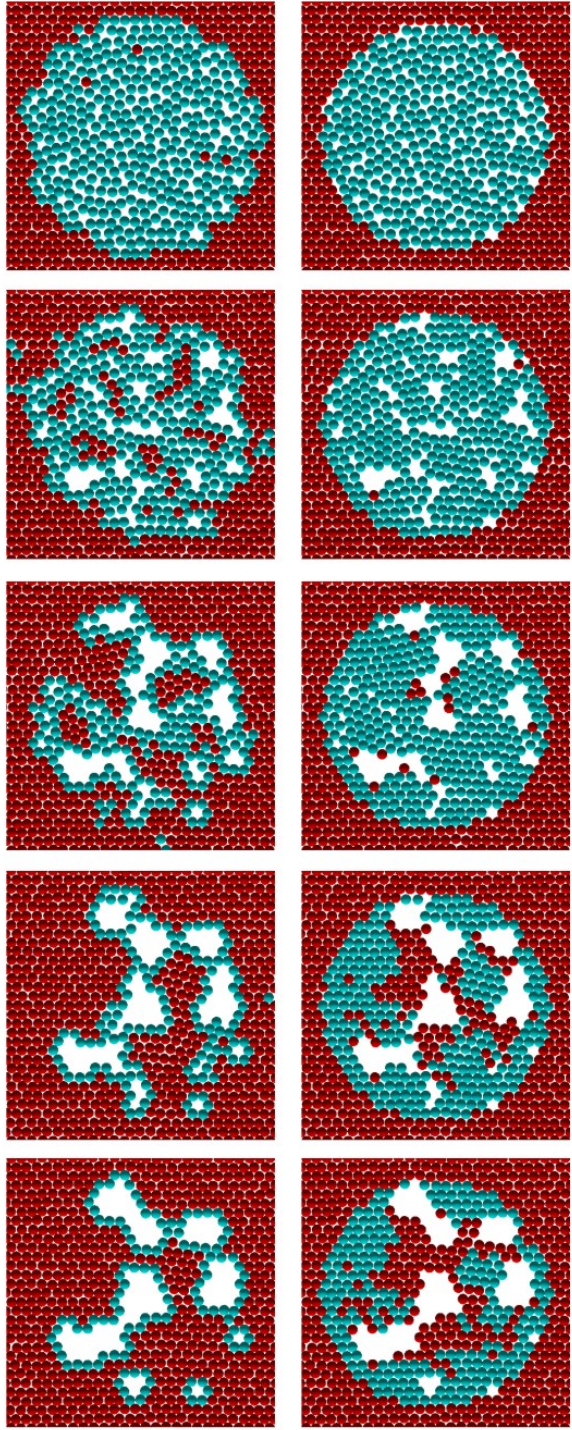

**Figure 7.** 2D crystallization of 350 particles (confined melt), PF = 0.72, T* = 0.12; Left – coloring according to the coordination number of each particle; Right – coloring according to each particle displacement against its initial position.

It is intriguing to compare the results presented on Figure 1 and Figure 7. Both simulations are at same packing fraction and temperature and the only difference is that on Figure 7 the liquid melt is confined inside an already crystalline phase. It is seen that in the latter case much more and bigger voids are formed in the final crystal structure. The striking difference between the two cases is clearly seen on the bottom images of right columns on Figs 1 and 7. In the case of free melt, the dominant movement of the particles is toward the center of the system, while for a confined melt the dominant movement is to the center of the system.

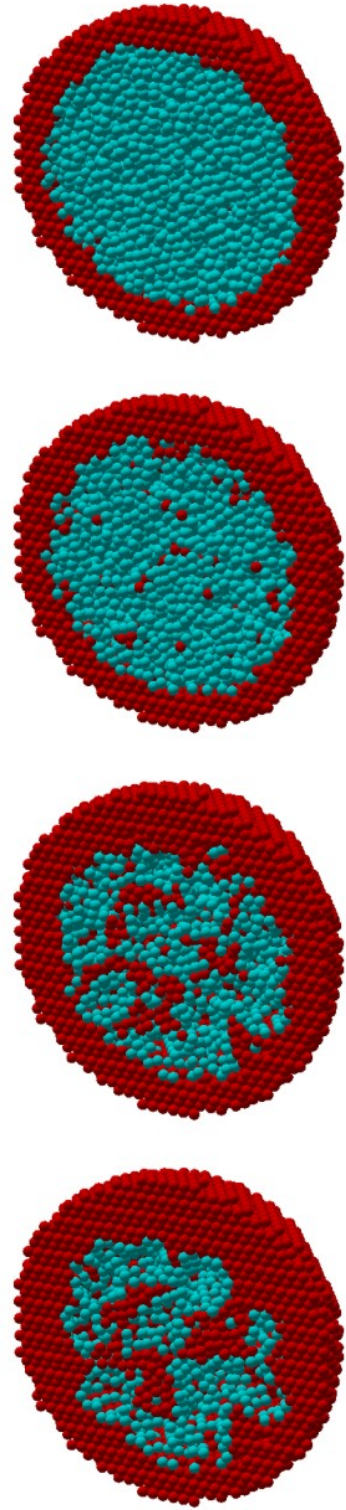

**Figure 8.** 3D crystallization of 7000 liquid particles confined in a *hcp* crystal core of 10000 particles, PF = 0.50, T* = 0.10. Only half of the system is shown (particles with positive value of z-coordinates).

This is the role of the already formed crystalline core – it prevents the shrinkage, thus the inner parts of the melt become a subject of stress accumulation and as a result the crystallization induced porosity is presented. Crystallization of a 3D liquid melt confined in an already formed (*hcp*) crystalline phase is shown on Figure 8. Only the outermost two layers of the crystalline phase are fixed. The melt crystallizes both in its volume and on the crystal core, perfectly reproducing the (*hcp*) arrangement. Thus the crystal core plays role of a sink for the melt phase particles, leading to a predominant consumption near the boundary melt/crystal core. Simultaneously into the volume crystal phase is also formed. This means that there are two "crystal fronts" which consume the liquid phase in a concurrent manner, and here one must have in mind the "focusing" effect of the curvature of the crystalline core (Karamanov A. et al., 2010). The whole system cannot shrinks since the outermost two surface layers are fixed, hence the liquid inside accumulates stress, and as a result (which may be thought as a stress relaxation (or stress responce) phenomenon) the liquid breaks forming a crystallization induced porosity.

## CONCLUSIONS

Monte Carlo simulations in 2D and 3D on a liquid melt that is a subject of crystallization using simple LJ potential show that the initial stage of crystallization starts with predominant forming of an overcritical nuclei into the sub-surface layer of the system. This leads to a formation of a rigid crystalline core that severely stops the volume shrinkage of the system, and as a result a considerable amount of liquid phase remains trapped into this core. As the crystallization of this liquid phase continues (second stage of the overall crystallization process), the liquid breaks and forms crystallization induced closed porosity. Nevertheless we use a simple short range interaction potential, our results shed light on the influence of the crystal core onto the crystallizing liquid melt. We show that CIP is a "natural" response effect of the system, due to the formation of a rigid crystalline "percolation" core.

BR is grateful for the financial help of Project BG051PO001-3.3.06-0038 funded by OP Human Resources Development 2007-2013 of EU Structural Funds.